\documentclass[aps,prl,reprint,superscriptaddress,amsmath, amssymb]{revtex4-1}

\usepackage{graphicx}
\usepackage[mathlines]{lineno}
\usepackage{amssymb}
\usepackage{amsmath}

\begin{document}

\preprint{APS/123-QED}

\title{Search for Cosmic-ray Boosted Sub-GeV Dark Matter using Recoil Protons at Super-Kamiokande}

\newcommand{\AFFicrr}{\affiliation{Kamioka Observatory, Institute for Cosmic Ray Research, University of Tokyo, Kamioka, Gifu 506-1205, Japan}}
\newcommand{\AFFkashiwa}{\affiliation{Research Center for Cosmic Neutrinos, Institute for Cosmic Ray Research, University of Tokyo, Kashiwa, Chiba 277-8582, Japan}}
\newcommand{\AFFicrronly}{\affiliation{Institute for Cosmic Ray Research, University of Tokyo, Kashiwa, Chiba 277-8582, Japan}}
\newcommand{\AFFipmu}{\affiliation{Kavli Institute for the Physics and
Mathematics of the Universe (WPI), The University of Tokyo Institutes for Advanced Study,
University of Tokyo, Kashiwa, Chiba 277-8583, Japan }}
\newcommand{\AFFmad}{\affiliation{Department of Theoretical Physics, University Autonoma Madrid, 28049 Madrid, Spain}}
\newcommand{\AFFubc}{\affiliation{Department of Physics and Astronomy, University of British Columbia, Vancouver, BC, V6T1Z4, Canada}}
\newcommand{\AFFbu}{\affiliation{Department of Physics, Boston University, Boston, MA 02215, USA}}
\newcommand{\AFFuci}{\affiliation{Department of Physics and Astronomy, University of California, Irvine, Irvine, CA 92697-4575, USA }}
\newcommand{\AFFcsu}{\affiliation{Department of Physics, California State University, Dominguez Hills, Carson, CA 90747, USA}}
\newcommand{\AFFcnm}{\affiliation{Institute for Universe and Elementary Particles, Chonnam National University, Gwangju 61186, Korea}}
\newcommand{\AFFduke}{\affiliation{Department of Physics, Duke University, Durham NC 27708, USA}}
\newcommand{\AFFfukuoka}{\affiliation{Junior College, Fukuoka Institute of Technology, Fukuoka, Fukuoka 811-0295, Japan}}
\newcommand{\AFFgifu}{\affiliation{Department of Physics, Gifu University, Gifu, Gifu 501-1193, Japan}}
\newcommand{\AFFgist}{\affiliation{GIST College, Gwangju Institute of Science and Technology, Gwangju 500-712, Korea}}
\newcommand{\AFFuh}{\affiliation{Department of Physics and Astronomy, University of Hawaii, Honolulu, HI 96822, USA}}
\newcommand{\AFFicl}{\affiliation{Department of Physics, Imperial College London , London, SW7 2AZ, United Kingdom }}
\newcommand{\AFFkek}{\affiliation{High Energy Accelerator Research Organization (KEK), Tsukuba, Ibaraki 305-0801, Japan }}
\newcommand{\AFFkobe}{\affiliation{Department of Physics, Kobe University, Kobe, Hyogo 657-8501, Japan}}
\newcommand{\AFFkyoto}{\affiliation{Department of Physics, Kyoto University, Kyoto, Kyoto 606-8502, Japan}}
\newcommand{\AFFliv}{\affiliation{Department of Physics, University of Liverpool, Liverpool, L69 7ZE, United Kingdom}}
\newcommand{\AFFmiyagi}{\affiliation{Department of Physics, Miyagi University of Education, Sendai, Miyagi 980-0845, Japan}}
\newcommand{\AFFnagoya}{\affiliation{Institute for Space-Earth Environmental Research, Nagoya University, Nagoya, Aichi 464-8602, Japan}}
\newcommand{\AFFkmi}{\affiliation{Kobayashi-Maskawa Institute for the Origin of Particles and the Universe, Nagoya University, Nagoya, Aichi 464-8602, Japan}}
\newcommand{\AFFpol}{\affiliation{National Centre For Nuclear Research, 02-093 Warsaw, Poland}}
\newcommand{\AFFsuny}{\affiliation{Department of Physics and Astronomy, State University of New York at Stony Brook, NY 11794-3800, USA}}
\newcommand{\AFFokayama}{\affiliation{Department of Physics, Okayama University, Okayama, Okayama 700-8530, Japan }}
\newcommand{\AFFosaka}{\affiliation{Department of Physics, Osaka University, Toyonaka, Osaka 560-0043, Japan}}
\newcommand{\AFFox}{\affiliation{Department of Physics, Oxford University, Oxford, OX1 3PU, United Kingdom}}
\newcommand{\AFFqmul}{\affiliation{School of Physics and Astronomy, Queen Mary University of London, London, E1 4NS, United Kingdom}}
\newcommand{\AFFregina}{\affiliation{Department of Physics, University of Regina, 3737 Wascana Parkway, Regina, SK, S4SOA2, Canada}}
\newcommand{\AFFseoul}{\affiliation{Department of Physics, Seoul National University, Seoul 151-742, Korea}}
\newcommand{\AFFsheff}{\affiliation{Department of Physics and Astronomy, University of Sheffield, S3 7RH, Sheffield, United Kingdom}}
\newcommand{\AFFshizuokasc}{\affiliation{Department of Informatics in
Social Welfare, Shizuoka University of Welfare, Yaizu, Shizuoka, 425-8611, Japan}}
\newcommand{\AFFstfc}{\affiliation{STFC, Rutherford Appleton Laboratory, Harwell Oxford, and Daresbury Laboratory, Warrington, OX11 0QX, United Kingdom}}
\newcommand{\AFFskk}{\affiliation{Department of Physics, Sungkyunkwan University, Suwon 440-746, Korea}}
\newcommand{\AFFtodai}{\affiliation{Department of Physics, University of Tokyo, Bunkyo, Tokyo 113-0033, Japan }}
\newcommand{\AFFtit}{\affiliation{Department of Physics,Tokyo Institute of Technology, Meguro, Tokyo 152-8551, Japan }}
\newcommand{\AFFtus}{\affiliation{Department of Physics, Faculty of Science and Technology, Tokyo University of Science, Noda, Chiba 278-8510, Japan }}
\newcommand{\AFFtoronto}{\affiliation{Department of Physics, University of Toronto, ON, M5S 1A7, Canada }}
\newcommand{\AFFtriumf}{\affiliation{TRIUMF, 4004 Wesbrook Mall, Vancouver, BC, V6T2A3, Canada }}
\newcommand{\AFFtokai}{\affiliation{Department of Physics, Tokai University, Hiratsuka, Kanagawa 259-1292, Japan}}
\newcommand{\AFFtsinghua}{\affiliation{Department of Engineering Physics, Tsinghua University, Beijing, 100084, China}}
\newcommand{\AFFynu}{\affiliation{Department of Physics, Yokohama National University, Yokohama, Kanagawa, 240-8501, Japan}}
\newcommand{\AFFllr}{\affiliation{Ecole Polytechnique, IN2P3-CNRS, Laboratoire Leprince-Ringuet, F-91120 Palaiseau, France }}
\newcommand{\AFFbari}{\affiliation{ Dipartimento Interuniversitario di Fisica, INFN Sezione di Bari and Universit\`a e Politecnico di Bari, I-70125, Bari, Italy}}
\newcommand{\AFFnapoli}{\affiliation{Dipartimento di Fisica, INFN Sezione di Napoli and Universit\`a di Napoli, I-80126, Napoli, Italy}}
\newcommand{\AFFroma}{\affiliation{INFN Sezione di Roma and Universit\`a di Roma ``La Sapienza'', I-00185, Roma, Italy}}
\newcommand{\AFFpadova}{\affiliation{Dipartimento di Fisica, INFN Sezione di Padova and Universit\`a di Padova, I-35131, Padova, Italy}}
\newcommand{\AFFkeio}{\affiliation{Department of Physics, Keio University, Yokohama, Kanagawa, 223-8522, Japan}}
\newcommand{\AFFwinnipeg}{\affiliation{Department of Physics, University of Winnipeg, MB R3J 3L8, Canada }}
\newcommand{\AFFkcl}{\affiliation{Department of Physics, King's College London, London, WC2R 2LS, UK }}
\newcommand{\AFFwarwick}{\affiliation{Department of Physics, University of Warwick, Coventry, CV4 7AL, UK }}
\newcommand{\AFFral}{\affiliation{Rutherford Appleton Laboratory, Harwell, Oxford, OX11 0QX, UK }}
\newcommand{\AFFwu}{\affiliation{Faculty of Physics, University of Warsaw, Warsaw, 02-093, Poland }}
\newcommand{\AFFbcit}{\affiliation{Department of Physics, British Columbia Institute of Technology, Burnaby, BC, V5G 3H2, Canada }}
\newcommand{\AFFtohoku}{\affiliation{Department of Physics, Faculty of Science, Tohoku University, Sendai, Miyagi, 980-8578, Japan }}
\newcommand{\AFFicise}{\affiliation{Institute For Interdisciplinary Research in Science and Education, ICISE, Quy Nhon, 55121, Vietnam }}
\newcommand{\AFFilance}{\affiliation{ILANCE, CNRS - University of Tokyo International Research Laboratory, Kashiwa, Chiba 277-8582, Japan}}
\newcommand{\AFFibs}{\affiliation{Institute for Basic Science (IBS), Daejeon, 34126, Korea}}

\AFFicrr
\AFFkashiwa
\AFFicrronly
\AFFmad
\AFFbu
\AFFbcit
\AFFuci
\AFFcsu
\AFFcnm
\AFFduke
\AFFllr
\AFFfukuoka
\AFFgifu
\AFFgist
\AFFuh
\AFFibs
\AFFicise
\AFFicl
\AFFbari
\AFFnapoli
\AFFpadova
\AFFroma
\AFFilance
\AFFkeio
\AFFkek
\AFFkcl
\AFFkobe
\AFFkyoto
\AFFliv
\AFFmiyagi
\AFFnagoya
\AFFkmi
\AFFpol
\AFFsuny
\AFFokayama
\AFFox
\AFFral
\AFFseoul
\AFFsheff
\AFFshizuokasc
\AFFstfc
\AFFskk
\AFFtohoku
\AFFtokai
\AFFtodai
\AFFipmu
\AFFtit
\AFFtus
\AFFtoronto
\AFFtriumf
\AFFtsinghua
\AFFwu
\AFFwarwick
\AFFwinnipeg
\AFFynu

\author{K.~Abe}
\AFFicrr
\AFFipmu
\author{Y.~Hayato}
\AFFicrr
\AFFipmu
\author{K.~Hiraide}
\AFFicrr
\AFFipmu
\author{K.~Ieki}
\author{M.~Ikeda}
\AFFicrr
\AFFipmu
\author{J.~Kameda}
\AFFicrr
\AFFipmu
\author{Y.~Kanemura}
\author{R.~Kaneshima}
\author{Y.~Kashiwagi}
\AFFicrr
\author{Y.~Kataoka}
\AFFicrr
\AFFipmu
% ...


\collaboration{The Super-Kamiokande Collaboration}
\noaffiliation

\date{\today}

\begin{abstract}

We report a search for cosmic-ray boosted dark matter with protons using the 0.37 megaton$\times$years data collected at Super-Kamiokande experiment during the 1996-2018 period (SKI-IV phase). 
We searched for an excess of proton recoils above the atmospheric neutrino background from the vicinity of the Galactic Center. 
No such excess is observed, and limits are calculated for two reference models of dark matter with either a constant interaction cross-section or through a scalar mediator.
This is the first experimental search for boosted dark matter with hadrons using directional information.
The results present the most stringent limits on cosmic-ray boosted dark matter and exclude the dark matter-nucleon elastic scattering cross-section between $10^{-33}\text{ cm}^{2}$ and $10^{-27}\text{ cm}^{2}$ for dark matter mass from 1 MeV/$c^2$ to 300 MeV/$c^2$.

\begin{description}
      \item[DOI]
   \end{description}
\end{abstract}

\maketitle

There is overwhelming evidence for the existence of dark matter~\cite{Zwicky:1933gu, Blumenthal:1984bp, Sofue:2000jx, Schumann:2019eaa, Bertone:2016nfn}.
The properties of the dark matter remain unknown beyond gravitational interaction, and there are a variety of theoretical models predicting a wide range of masses for dark matter candidates (e.g.~\cite{Essig:2011nj, Knapen:2017xzo, Smirnov:2019ngs}).
Despite significant efforts of highly sensitive direct dark matter detection experiments to probe interactions of dark matter at the mass range of GeV/$c^2$ to TeV/$c^2$, dark matter have been elusive thus far~\cite{LUX:2016ggv, XENON:2018voc}. 
Meanwhile, dark matter at the sub-GeV mass range is poorly explored~\cite{Feng:2008ya, Boehm:2003hm, Lin:2011gj, Hochberg:2014kqa}.
The conventional direct dark matter detection experiments focusing on nuclear recoils are not sensitive to cold sub-GeV dark matter due to insufficient recoil energy, and the experimental searches of cold sub-GeV dark matter have focused on the Migdal effect~\cite{Ibe:2017yqa, PhysRevLett.123.241803, PhysRevLett.123.161301, PhysRevD.99.082003} and the interaction with electrons~\cite{Essig:2011nj, SENSEI:2020dpa, EDELWEISS:2020fxc}.
Besides, if a fraction of the cold dark matter is boosted to relativistic energies, it can be efficiently detected in direct detection experiments as well as higher threshold neutrino detectors~\cite{Agashe:2014yua, PhysRevD.95.075018, PhysRevD.97.063007, Hu:2016xas, Giudice:2017zke, Cappiello:2019qsw, Arguelles:2022fqq}.

A general possibility for dark matter to obtain relativistic energies is via the upscattering by cosmic-rays, constituting cosmic-ray boosted dark matter (CRDM)~\cite{Bringmann:2018cvk, Ema:2018bih, Cappiello:2018hsu, Cappiello:2019qsw}.
The upscattering process originates from the same dark matter-nucleus interactions as direct detection experiments search for, without requiring additional assumptions or model dependence.
Due to the dark matter density distribution concentrated toward the Galactic Center (GC)~\cite{Navarro:1995iw}, the CRDM arriving at the Earth has a directional preference from the GC.
For terrestrial detectors, the CRDM-nucleon interaction in the Earth can be sizable, and the dark matter can be scattered multiple times and become attenuated when traveling through the Earth~\cite{PhysRevLett.126.091804}.

The boosted relativistic component can be observed by the interactions in the detector with electrons~\cite{Agashe:2014yua, Ema:2018bih} or hadrons~\cite{Ema:2020ulo, Cappiello:2019qsw}.
In 2018, the Super-Kamiokande experiment published the first experimental search for boosted dark matter in a terrestrial detector with electron recoils~\cite{Super-Kamiokande:2017dch}.
Later on, PROSPECT~\cite{PhysRevD.104.012009}, PandaX-II~\cite{PandaX-II:2021kai}, and CDEX-10~\cite{CDEX:2022fig} reported their result on CRDM using nuclear recoils, setting cross-section limits at $10^{-31}-10^{-26}$~cm$^2$ in a dark matter mass region from MeV/$c^2$ to GeV/$c^2$.

In this analysis, we search for CRDM from MeV/$c^2$ to GeV/$c^2$ with recoil protons at the Super-Kamiokande (SK) experiment~\cite{Super-Kamiokande:2002weg}.
We use the data collected at SK during the 1996-2018 period (SKI-IV phases).
The large fiducial volume and the directional reconstruction ability of SK, a water Cherenkov detector, enables a sensitive search for CRDM.
The parameter space we explore extends by more than one order of magnitude beyond the existing limits~\cite{PhysRevD.104.012009, PandaX-II:2021kai}.

Super-Kamiokande is a cylindrical 50~kiloton water Cherenkov detector located in Kamioka, Japan, under a 2,700~meter water-equivalent rock overburden~\cite{Super-Kamiokande:2002weg}.
The detector consists of an inner detector (ID) and an outer detector (OD) optically separated at 2~m from the detector's outer wall. 
There are 11,129 inward-facing 20-inch PMTs viewing the 32~kton target volume of the ID, and the OD is viewed by 1,885 outward-facing 8-inch PMTs.
The ID is used to reconstruct the energies, vertices, and to perform the particle identifications (PID) of the physics events, while the OD is primarily used as a veto for charged particles entering from outside the detector or identifying particles that exit the ID.

This analysis uses the fully contained fiducial volume (FCFV) dataset composed of events that have activity only in the ID (FC) and are reconstructed with vertices more than 2~m from the ID wall, corresponding to the 22.5~kton fiducial volume (FV).
The total livetime of the dataset is 6050.3~days, corresponding to an exposure of 0.37~megaton$\times$years.
The visible energy, corresponding to the energy of an electron that would cause the same amount of light in the detector, of the events is required to be above 30~MeV to remove spallation backgrounds induced by the cosmic-ray muons.
To select recoil protons without extra activities, we require the candidate events to have only one single reconstructed Cherenkov ring.

In this FCFV sample, the majority of events are electrons and muons.
Electrons create electromagnetic showers which produce fuzzy rings and can be easily removed, while muons and protons have a sharp ring edge.
To select proton events from the muon background, we employed a proton fitter that utilizes the light pattern and ring topology to calculate the proton likelihood, proton momentum, and track length~\cite{Super-Kamiokande:2009kfy}.
A distinctive feature of the protons is that they are likely to have hadronic interactions in water and lose energy by producing secondary particles.
If both the secondary particles and the scattered proton are below Cherenkov threshold, the Cherenkov light emission is truncated and leaves a narrow proton Cherenkov ring.
If the secondary particles, typically pions, are energetic enough to emit bright Cherenkov light, the identification of the proton becomes significantly more difficult, and therefore the reconstruction is less efficient for higher momentum protons due to the higher hadronic interaction probability.

Since the identification performance depends on proton momentum, we established a series of kinematic precuts.
To remove the majority of high energy muons, we require the reconstructed proton momentum to be less than 3~GeV/$c$, the visible energy to be less than 400~MeV, and the corrected charge within $70^\circ$ of the direction~\cite{refmat} to be less than 2,000 photo-electrons.
Due to the large mass, protons have a smaller Cherenkov angle compared to muons at the same momentum, and thus we require the reconstructed Cherenkov angle of candidate events to be less than 40$^\circ$.
Finally, we place a cut on the proton-muon identification likelihood.

To further enhance the proton-muon separation, a multi-variate analysis (MVA) is employed after the precuts.
The input variables include the fitted track length, the fitted momentum, and the PID likelihood from the proton fitter~\cite{Super-Kamiokande:2009kfy}, the charge distribution within and outside of the Cherenkov ring, the reconstructed Cherenkov angle, the vertex reconstruction quality, and the number of decay-electrons.
More details on the variable definitions and distributions can be found in the supplementary material~\cite{refmat}.

\begin{figure}
	\centering
			\includegraphics[width=1.00\linewidth]{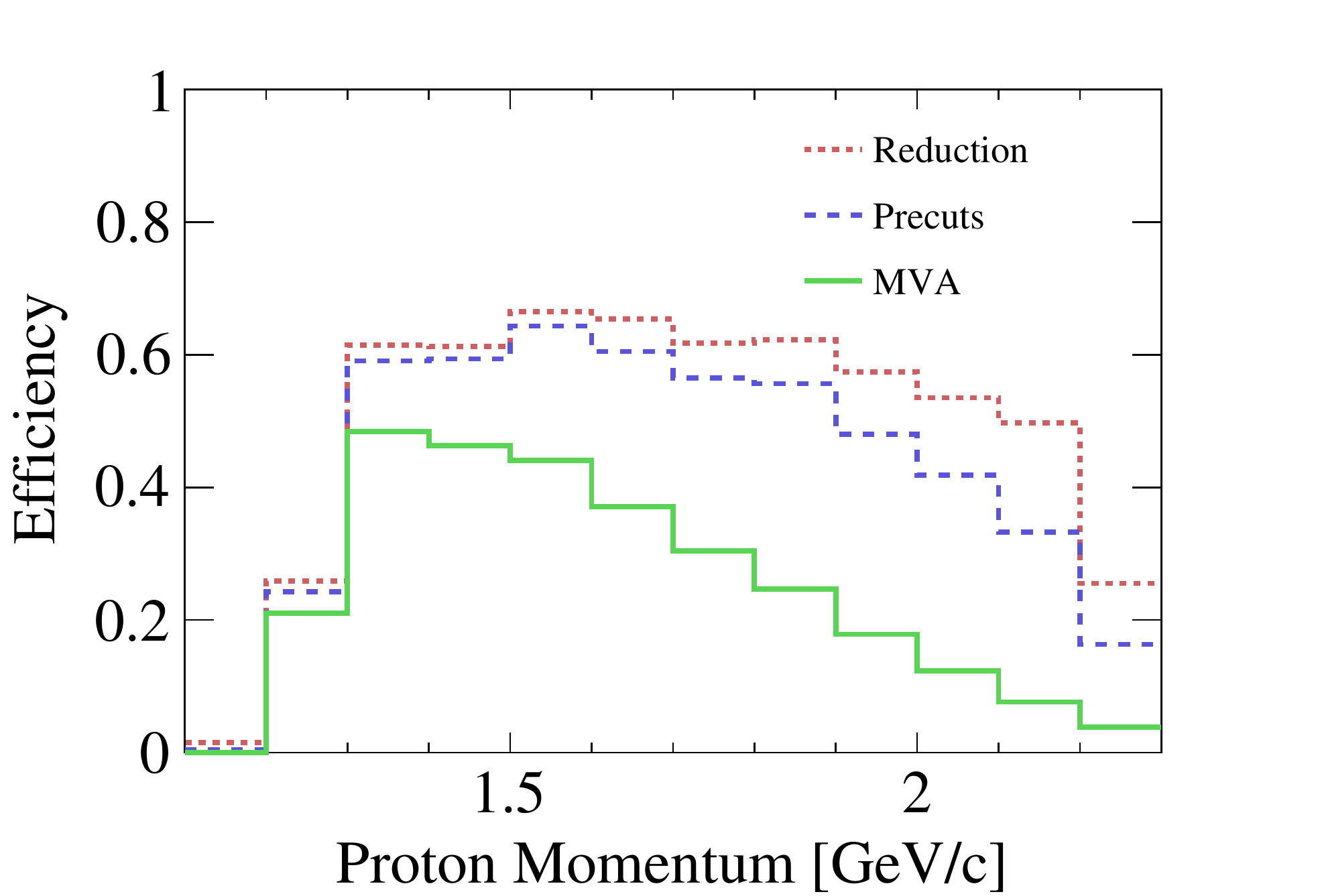}
			\caption{The selection efficiencies for the proton sample. The red dotted line indicates the reduction efficiency of the FCFV sample above 30~MeV. The blue dashed line represents the efficiency after precuts. The green solid line is the efficiency after the MVA cut.}
			\label{fig:protontotaleff}
\end{figure}

The structure of the MVA is selected as a multilayer perceptron~\cite{Hocker:2007ht}, which is trained with simulated protons and non-proton events from the atmospheric neutrino MC sample after the precuts.
The MVA takes the eight input variables and outputs an estimator describing how signal- or background-like an event is.
The cut on the MVA estimator is optimized towards best sensitivity assuming a 0.37~megaton$\times$years exposure and realistic systematic errors, and the corresponding efficiency is shown in Fig.~\ref{fig:protontotaleff}.
The proton reconstruction is only feasible within a momentum window between 1.2~GeV/$c$ and 2.3~GeV/$c$.
Below 1.2~GeV/$c$, the Cherenkov light yield is too low to reconstruct the proton ring.
Above 2.3~GeV/$c$, the protons tend to have hadronic interactions and the secondary particles make extra rings, which complicates the proton reconstruction.
After the precuts and the MVA cut, we expected 86.0 proton events and 25.7 non-proton events in the final sample from atmospheric neutrinos.

The systematic uncertainties in this proton sample include uncertainties in atmospheric neutrino cross-section and flux (26\%), proton hadronic interaction systematics (4\%), and detector related systematics (8\% for proton events, and 13\% for non-proton background events).
The major source of the atmospheric neutrino related uncertainty is the neutral current / charged current ratio (20\%).
In summary, we estimated 27\% for protons from atmospheric neutrinos, 29\% uncertainty for non-proton background events from atmospheric neutrinos, and 9\% in proton signal efficiency.
As such, we expected $111.7\pm10.6\text{(stat.)}\pm30.7\text{(sys.)}$ events for the searched 0.37 megaton$\times$years livetime in the final sample from atmospheric neutrinos.
Compared with the observation of 126 events, this result is within the estimated systematic and statistical uncertainty.

The CRDM flux is determined by the dark matter distribution model, the cosmic-ray model, and the dark matter interaction model.
In this analysis, we use the NFW profile for Galactic dark matter density distribution~\cite{Navarro:1995iw}. 
For simplicity, the cosmic-ray flux is assumed to be homogeneous within a leaky box model cylinder~\cite{Strong:2007nh}, and the radius and height of the cylinder are taken as $R=10$~kpc and $h=1$~kpc following Ref.~\cite{Bringmann:2018cvk, Ema:2020ulo}.
The energy spectrum of cosmic-rays is modeled from 10~MeV to above 50~GeV with Voyager's observation~\cite{Cummings:2016pdr} and different theoretical calculations~\cite{Boschini:2017fxq, Tomassetti:2019tyc}, as specified in Ref.~\cite{Ema:2020ulo}.
For the dark matter nucleon interaction cross-section, we consider two reference scenarios, one with fermionic dark matter and a scalar mediator, and one with a constant dark matter-nucleon interaction cross-section.
In the scalar mediator scenario, we employed the flux and cross-section as calculated in Ref.~\cite{Ema:2020ulo} with a mediator mass of $m = 1$~GeV/$c^2$.
For the constant cross-section dark matter model, we make use of a reproduced flux from Ref.~\cite{Bringmann:2018cvk}, and the cross-section is assumed to be $10^{-30}$~cm$^2$ at the dark matter-nucleon coupling constant $g=1$.

As SK is a Cherenkov detector, it can reconstruct directions of the recoil protons, which facilitates the separation of the relatively isotropic atmospheric neutrino backgrounds from signals that are more peaked in the direction of the GC.
The directional distribution of recoil protons with regard to the GC is a convolution of the angular resolution of proton rings, the kinematic correlation between recoil proton direction and the incoming CRDM, and the model-dependent directional distribution of the CRDM flux.
The reconstructed angular resolution of proton rings is 2.6$^\circ$, a subdominant factor compared to the kinematic angular correlation and the CRDM distribution. 
Considering the two reference cross-section models and the different cylinder sizes for cosmic-ray modeling, we found that the optimal directional cuts from the GC varies by about $10\%$. 
For a more general interpretation, we fix the GC direction cut at $\cos\theta_{GC}>0.6$.

At the large dark matter coupling scale we are probing, the CRDM attenuation within the Earth is non-negligible, which ensures that the CRDM flux arriving at the detector comes primarily from above the horizon.
To reject the upward-going atmospheric neutrino backgrounds and to avoid the uncertainty near the horizon, we apply a zenith angle cut at $-\cos\theta_z>0.2$.
The efficiency for such a cut can be obtained by calculating the fraction of live-time the GC is above the horizon considering the latitude of the observatory site, which is $0.29$ for SK.
After the GC direction cut and the zenith angle cut, the expected number of backgrounds from atmospheric neutrinos in the proton sample is expected to be 7.4 (6.5) events with (without) normalization to data.

\begin{figure}
   \centering
         \includegraphics[width=1.00\linewidth]{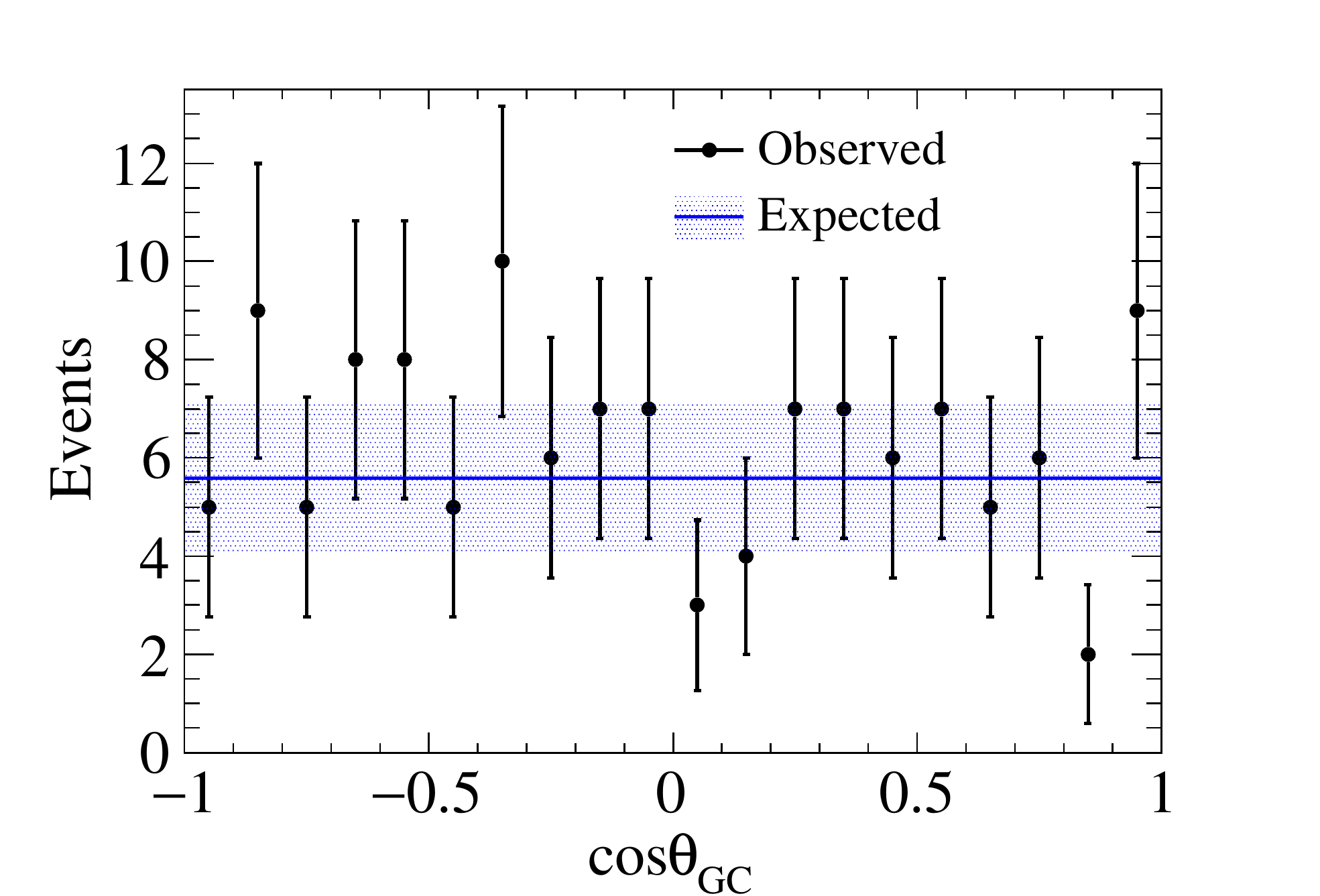}
         \includegraphics[width=1.00\linewidth]{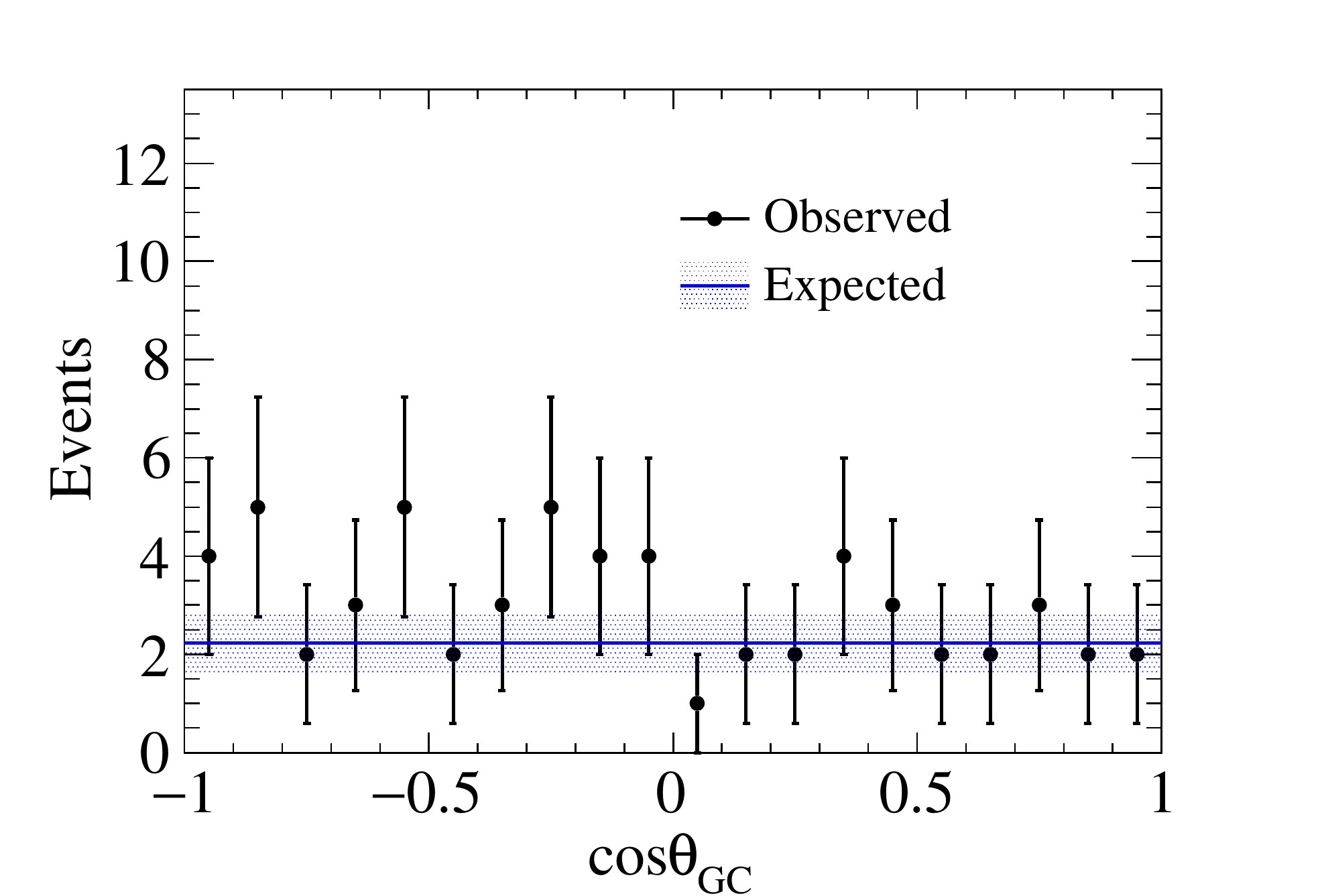}
         \caption{The angle between proton ring and the GC for events in the proton sample, without (upper) and with (lower) the zenith angle cut. The black points indicate data with statistical uncertainty. The blue bands indicate MC expectation with systematic uncertainty.}
         \label{fig:UDMGCdir}
\end{figure}

The GC angular distribution of the MC expectation and data with and without the zenith cut are shown in Fig.~\ref{fig:UDMGCdir}.
To avoid the systematic bias from the atmospheric neutrino azimuthal spectra, we employed an on-off source search, with the on-source at the GC, and the off-source shifted from the on-source by 180$^\circ$ in right ascension, as shown in the supplementary material~\cite{refmat}.
Applying the cut $-\cos\theta_z>0.2$ and $\cos\theta_{GC}>0.6$, the remaining number of events in the proton data sample is 9 for the on-source (GC), and 7 for the off-source.
Considering the systematic uncertainty, the upper limit on the number of the CRDM recoil proton events can be calculated using Rolke method~\cite{Rolke:2004mj} as 5.7 events at 90\% confidence level.

\begin{figure}
   \centering
         \includegraphics[width=1.00\linewidth]{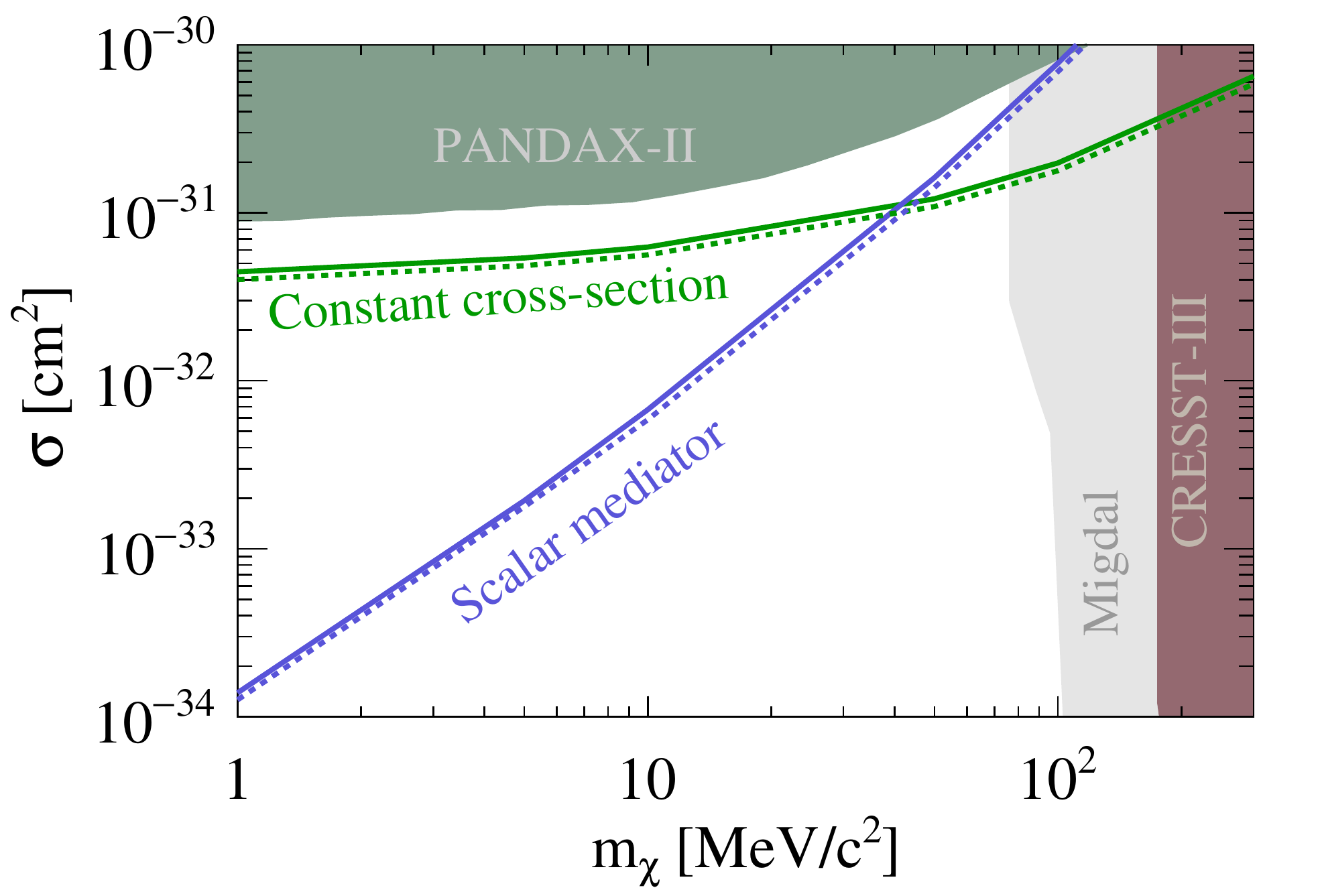}
			\caption{Constraints on dark matter-nucleon cross-section.
Solid lines show the upper limit while dashed lines indicate the sensitivity. 
The green lines are calculated with a constant cross-section model.
The blue lines are the cross-sections at the non-relativistic limit ($\sigma_{NR}$) for scalar mediator model.
The shaded sage green region indicates the PANDAX-II CRDM exclusion region~\cite{PandaX-II:2021kai}.
	The shaded maroon region shows the CRESST-III exclusion region~\cite{CRESST:2019jnq} and the shaded grey region shows the constraints via Migdal effect from CDEX-1B~\cite{CDEX:2019hzn} and XENON1T~\cite{XENON:2019zpr}.}
         \label{fig:UDMcross}
\end{figure}

In the absence of an excess in the proton sample, we calculated the upper limit of the dark matter-nucleon coupling and the interaction cross-section.
Note that the CRDM is produced from the same mechanism of dark matter-nucleon scattering, and therefore the CRDM flux is also proportional to the cross-section.
Our result covers the sub-GeV dark matter mass from MeV/$c^2$ to GeV/$c^2$ at~$10^{-33}$~cm$^2$, as shown in Fig.~\ref{fig:UDMcross}.

The recent CRDM search result from PANDAX-II~\cite{PandaX-II:2021kai} assuming constant cross-section is also shown for comparison. Due to the large exposure of SK and the directional information from the Cherenkov ring, the constraint from SK is better than the existing limits by a factor of 2.

If the dark matter-nucleon coupling is large enough, the CRDM flux will lose energy when traveling through the rock overburden above the detector, imposing an upper bound on the exclusion region.
This energy can be calculated with an analytical approximation considering the nuclear form factor effect~\cite{Xia:2021vbz}.
In the case of SK, due to the higher detection threshold from proton Cherenkov radiation, the experiment is only sensitive to sub-GeV dark matter above 0.5~GeV kinematic energy, and the attenuation of the rock overburden for this energy range is calculated to be below 10\% at $\sigma<10^{-27}$~cm$^2$.
Above $10^{-27}$~cm$^2$, the parameter space has been excluded by an analysis using cosmic microwave background data~\cite{Xu:2018efh}.
The lower end of the search range in dark matter mass at 1~MeV/$c^2$ is constrained by the Big Bang nucleosynthesis~\cite{Reno:1987qw, PhysRevD.101.123022}.
At higher dark matter mass, the constraints mainly come from the direct detection experiment CRESST-III~\cite{CRESST:2019jnq} and the Migdal effect searches at CDEX-1B~\cite{CDEX:2019hzn} and XENON1T~\cite{XENON:2019zpr}.

In summary, we report a directional search for the CRDM using a newly constructed proton sample selected from the data collected at Super-Kamiokande during the period of 1996-2018 (SKI-IV phases).
In the absence of an excess from dark matter signals above the expected background, we derived new limits on the dark matter-nucleon interaction cross-section, which are the most stringent constraint on hadronic coupling of sub-GeV dark matter so far.
This result benefits from the large fiducial volume and directional reconstruction ability of SK, which motivates further exploration of CRDM and boosted dark matter in general from the next generation large neutrino detectors with directional capabilities, such as Hyper-Kamiokande~\cite{Abe:2011ts} and DUNE~\cite{DUNE:2020ypp}.
The reported proton sample efficiency and direction distribution can also be interpreted by any theory that predicts an excess of proton recoils from the direction of the GC.


We thank Dr. Yohei Ema for providing the CRDM flux and insightful discussions.
We gratefully acknowledge the cooperation of the Kamioka Mining and Smelting Company.
The Super-Kamiokande experiment has been built and operated from funding by the 
Japanese Ministry of Education, Culture, Sports, Science and Technology, the U.S.
Department of Energy, and the U.S. National Science Foundation. Some of us have been 
supported by funds from the National Research Foundation of Korea NRF‐2009‐0083526
(KNRC) funded by the Ministry of Science, ICT, and Future Planning and the Ministry of
Education (2018R1D1A3B07050696, 2018R1D1A1B07049158), 
the Japan Society for the Promotion of Science, the National
Natural Science Foundation of China under Grants No. 11620101004, the Spanish Ministry of Science, 
Universities and Innovation (grant PGC2018-099388-B-I00), the Natural Sciences and 
Engineering Research Council (NSERC) of Canada, the Scinet and Westgrid consortia of
Compute Canada, the National Science Centre, Poland (2015/18/E/ST2/00758),
the Science and Technology Facilities Council (STFC) and GridPPP, UK, the European Union's 
Horizon 2020 Research and Innovation Programme under the Marie Sklodowska-Curie grant
agreement no.754496, H2020-MSCA-RISE-2018 JENNIFER2 grant agreement no.822070, and 
H2020-MSCA-RISE-2019 SK2HK grant agreement no. 872549.

\bibliography{CRDM}

\end{document}